\documentstyle[aps,epsfig]{revtex}
\def\be{\begin{equation}}
\def\ee{\end{equation}}
\begin{document}
 \title{Entanglement entropy with interface defects}
 \author{
Ingo Peschel\\
{\small Fachbereich Physik, Freie Universit\"at Berlin,} \\
{\small Arnimallee 14, D-14195 Berlin, Germany}}
 \maketitle
 \begin{abstract}
 We consider a section of a half-filled chain of free electrons and its
 entanglement with the rest of the system in the presence of one or 
 two interface defects. We find a logarithmic behaviour of the 
 entanglement entropy with constants depending continuously on 
 the defect strength.

\end{abstract}

\vspace{1cm}

The entanglement of different parts of a quantum system in the total
wave function can be viewed as the result of a coupling across their
common interface. Correspondingly, the entanglement entropy turns
out to be proportional to the interface area in simple models 
\cite{Bom86,Sred93,Plenio04}.
Modifications of the interface will therefore modify the entanglement
and in one-dimensional systems with short-range interactions, where the 
interface reduces to a point, a single defect is expected to have a marked
influence. This was recently pointed out by Levine \cite{Levine04} who
applied this idea to the case of a Luttinger liquid with one impurity. He
obtained within 
perturbation theory a reduction of the entanglement entropy $S$ which is 
strong for repulsive and weak for attractive interactions, if one considers 
a large subsystem. Roughly speaking, this corresponds to the known influence
of interactions on the impurity strength in this case \cite{Kane/Fisher92}.
Levine's approach using bosonization is interesting, but it also suggests 
to look directly at the simplest case and to investigate the problem for 
non-interacting electrons with specific defects.\\

In the present paper we therefore study free electrons hopping on a chain
for the case of half filling. In spin language, this corresponds to an XX model.
This is a critical system and the entanglement entropy between a subsystem
of length $L$ and the rest is given by the conformal result 
\cite{Holzhey94,Calabrese04,Vidal03,Latorre03,Jin04}
\be 
  S= \frac{c} {3} \ln L + k
  \label{eqn:entro1}
\ee
with the value $c=1$ for the central charge. How does this change in the
presence of a defect at the chosen interface ? For one case, the answer is
known : if the defect is such that it cuts the chain, the subsystem has a free
end and thus only one connection to the rest remains. This changes $c \rightarrow
c_{eff}=c/2$ and $k \rightarrow k'$, but the logarithm is unaffected \cite{Calabrese04}.
In the following we show that the same holds for arbitrary defects, either at
one or at both interfaces. These defects can be changed bonds or changed
site energies. The calculations are numerical and based on the determination
of the reduced density matrix $\rho$ from the one-particle correlation function
of the total chain \cite{Peschel03,Vidal03,Cheong/Henley02}. The necessary 
formulae are given in Section I. Eigenvalue spectra of $\rho$ for various defects, 
which differ in a characteristic way from the spectrum of the homogeneous system, 
are presented in Section II. From the eigenvalues, the entanglement entropy 
$S= -tr(\rho \ln \rho)$ is obtained and discussed in Section III. We determine the 
effective central charge $c_{eff}$ for bond and site defects and discuss its 
behaviour in terms of a simple model. In contrast to the constant $k$, it is always 
reduced by a defect and therefore the entanglement always becomes smaller if 
the size $L$ of the subsystem is large enough. A brief summary is given in Section IV.

\section{Basic formulae}

We consider  a system of free fermions hopping between neighbouring 
sites of an infinite linear chain. The corresponding Hamiltonian reads \\
 \be
  \hat{H}=- \sum_{n}  t_n (c_n^{\dagger} c_{n+1} + c_{n+1}^{\dagger} c_{n})
  + \sum_{n} \Delta_n c_n^{\dagger}c_{n} 
  \label{eqn:hop}
 \ee
where 
$t_n$ is the hopping matrix element and the 'hat' denotes quantities of the 
total system. In the following, we set $t_n =1 $ and $\Delta_n = 0$ except at the boundaries 
of the subsystem which consists of the sites $i = 1, 2, ...L$. We will mainly consider the case
of one bond defect at the left boundary, $t_0 = t$, but we will also give results for
two equal bond defects at the two boundaries, $t_0 = t_L= t$, 
and one or two site defects next to the boundary,
$\Delta_1 = \Delta$ or $\Delta_1 = \Delta_L = \Delta$.\\

The total system is assumed to be half filled and in its ground state $|0>$. The reduced density
matrix then has the form \cite{Peschel03,Peschel04}
\be
   \rho= {\cal K} \exp{(-\sum_{i,j} H_{ij} c_i^{\dagger} c_j )}
   \label{eqn:rho1}
 \ee
 where $\cal K$ is a normalization constant and
 the matrix $H_{ij}$ follows from the one-particle correlation function
 of the total system
 \be
  \hat{C}_{mn} = <0|\; c_m^{\dagger} c_n\; |0>
  \label{eqn:cf}
 \ee
 via the relation
 \be
    H = \ln{\,[(1-C)/C\,]}
  \label{eqn:HC}
 \ee
 Here $C$ denotes the $L \times L$ submatrix of $\hat{C}$ with 
 the sites restricted to the subsystem. For a homogeneous infinite system,
 $C$ is given by
 \be
  C_{ij} = C^{0}(i-j) =  \frac{\sin \bigl[\frac{\pi} {2}\,(i-j)\bigr]} {{\pi}\,(i-j)} 
  \label{eqn:cf1}
 \ee
 With defects, the translational invariance is lost and for single defects  
 $C$ has the general form
 \be
  C_{ij} = C^{0}(i-j) -  C^{1}(i+j)
  \label{eqn:cf2}
 \ee
 Thus $C_{ij}$ is the difference of a Toeplitz matrix depending on $(i-j)$
 and a Hankel matrix depending on $(i+j)$.
 Physically, the term $C^{1}$ leads to oscillatory behaviour of the correlations
 as one moves along the chain. For site defects, these are the Ruderman-
 Kittel oscillations if one looks at the density $C_{ii}$. For bond defects, the
 density is unaffected but the oscillations appear e.g. in the nearest-neighbour
 (bond) correlation function $C_{i,i+1}$. The quantity  $C^{1}$ can be obtained
 by using the scattering phase shifts and including the contribution of possible 
 localized states caused by the defect. Thus a single weak bond 
 $t= e^{\:\nu} \leq 1$ in an infinite total system leads to
 \be
      C^{1}(l) = - \frac {1} {2} \sinh \nu \; (e^{-\nu} I_l - e^{\nu} I_{l-2})
   \label{eqn:cf3}
  \ee
  where
 \be
   I_l = \int_{0}^{\pi /2} \frac {dq} {\pi} \frac {\cos(ql)} {\sinh^2 \nu + \sin^2 q}
   \label{eqn:int}
  \ee
 For $t=0$, this simplifies to $C^{1}(l)= C^{0}(l)$ and $C$ assumes the known form 
 for a system with an open end. For a strong bond, $t > 1$, one has to add the
 contribution
 \be 
  C^{1}(l) =- \sinh \nu \; \exp{(-\nu (l-1))}
  \label{eqn:boundstate}
 \ee
 from the occupied bound state at energy $-2\cosh \nu$ below the band.
 Similarly, a site defect $\Delta = 2 \sinh \nu >  0$ leads to
 \be
 C^{1}(l) =  \frac {1} {2} \sinh \nu \; ( I_{l-1} - I_{l-3}  +  2  \sinh \nu \; I_{l-2})
 \label{eqn:cf4}
  \ee
 By diagonalizing $C$ one obtains its eigenvalues $\zeta_k$  
 ($0 <\zeta_k <1$),  from which the eigenvalues $\varepsilon_k = \ln ((1-\zeta_k)/\zeta_k)$ 
 of $H$ ($-\infty < \varepsilon_k < \infty$) follow. The reduced density 
 matrix then takes the diagonal form
 \be
   \rho = {\cal K} \exp{(-\sum_{k} \varepsilon_ k c_k^{\dagger} c_k)}
   \label{eqn:rho2}
  \ee
  The entanglement entropy is defined by $S= -tr(\rho \ln \rho)$ (we are using the
  natural logarithm, not the one to basis 2) and reads
  in terms of the $\zeta_k$
  \be
   S = -\sum_{k} \bigl[\,\zeta_k \ln \zeta_k + (1-\zeta_k) \ln(1-\zeta_k) \bigr]
   \label{eqn:entro2a}
  \ee
  and in terms of the $\varepsilon_k$
 \be
  S= \sum_{k} \left [ \, \ln{[1+exp{(-\varepsilon_k)}]} +  \frac {\varepsilon_k}
        {\exp{(\varepsilon_k)}+1} \right ]
  \label{eqn:entro2b}
 \ee
    
 Only the $\zeta_k$ which are not too close to 0 or 1 resp. the $\varepsilon_k$
 which are not much larger than 1 in magnitude give a sizeable contribution to the
 entropy.
 
\section{Density-matrix spectra}

We have calculated the correlation matrix $C$ for the case of single defects  
by evaluating the integrals for $C^{1}$ numerically. From this the low-lying
single-particle eigenvalues $\varepsilon_k$ with  $ |\varepsilon_k| \leq 25$
were obtained. Higher values cannot be reached with standard double precision
routines because the corresponding $\zeta_k$ lie too close to 0 or 1. However, as
noted above, they are unimportant for $S$. In the following
we always consider even $L$.\\

Spectra for one bond defect are shown in Fig. 1 for a subsystem of $L=50$ sites.
The eigenvalues come in pairs ($\varepsilon,-\varepsilon$) due to the particle-hole
symmetry of the problem.
For a homogeneous system, $t=1$, one finds the slightly bent curve known
from previous investigations \cite{Chung/Peschel01,Cheong/Henley03}.\\

For a weak defect, one can see a shift of the dispersion curve which is upward
for positive $\varepsilon_k$ and downward for negative ones. This shift
becomes stronger as the bond becomes weaker. In addition, oscillations appear which
also increase for weaker bonds. Most importantly, however, more and more of the low
$\varepsilon_k$ follow a steeper dispersion curve as $t$ approaches zero. This
curve has a slope approximately twice as large as for the homogeneous system and no wiggles 
and represents the result one obtains directly for the system with a free end.
Thus the spectrum with the defect is a mixture of those for the two cases $t=1$ and $t = 0$.\\

For a strong defect, the situation is basically the same. However, here the two lowest eigenvalues 
$\varepsilon_0$ and $\varepsilon_1$ play a special role because they approach zero as 
$t\rightarrow \infty$ . This moves the left and right part of the spectra two units apart. 
This feature can be understood from the correlations in this limit. The two

\pagebreak
 
\begin{figure}
\centerline{\psfig{file=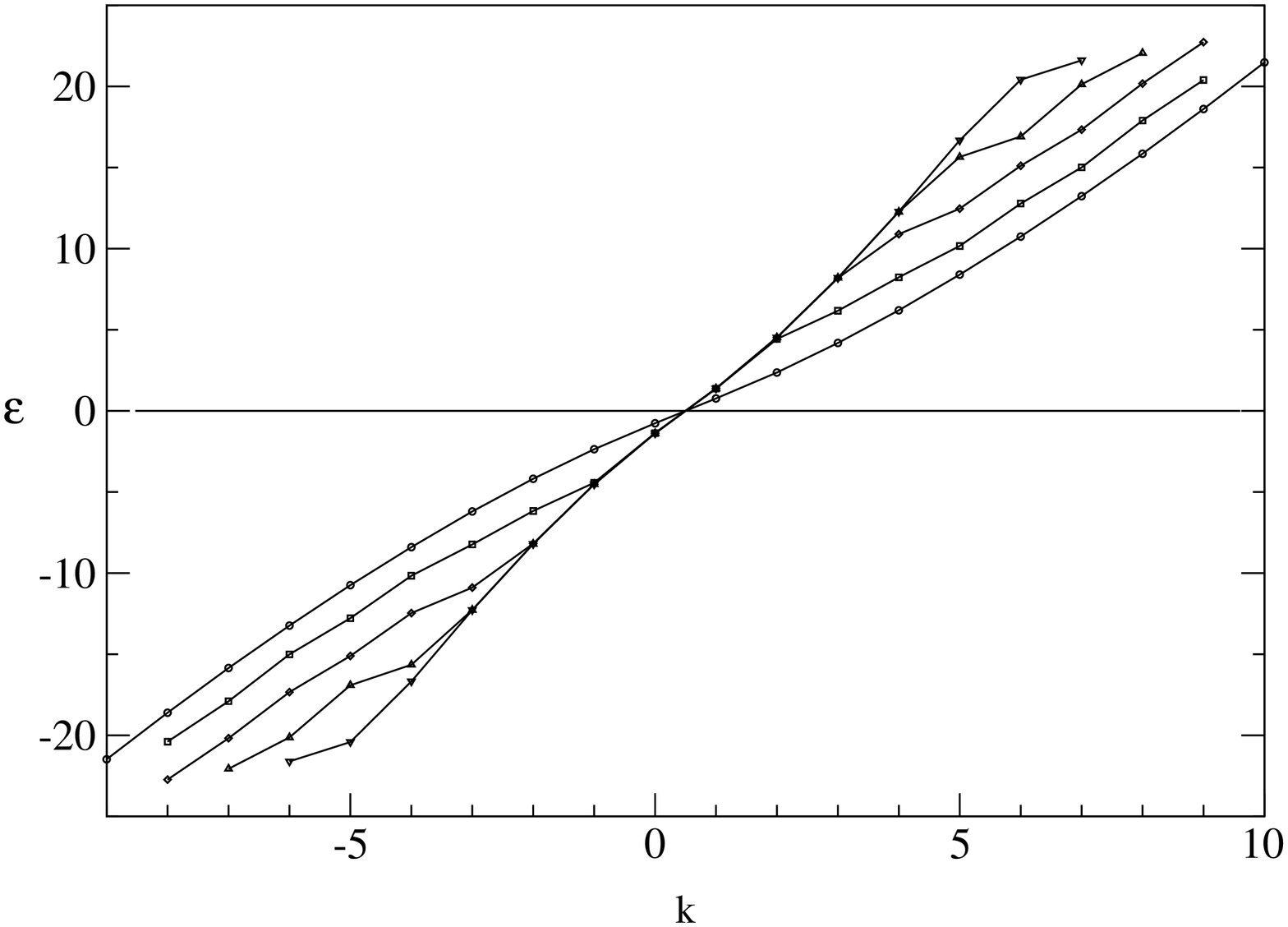,width=7cm,angle=-90}
\psfig{file=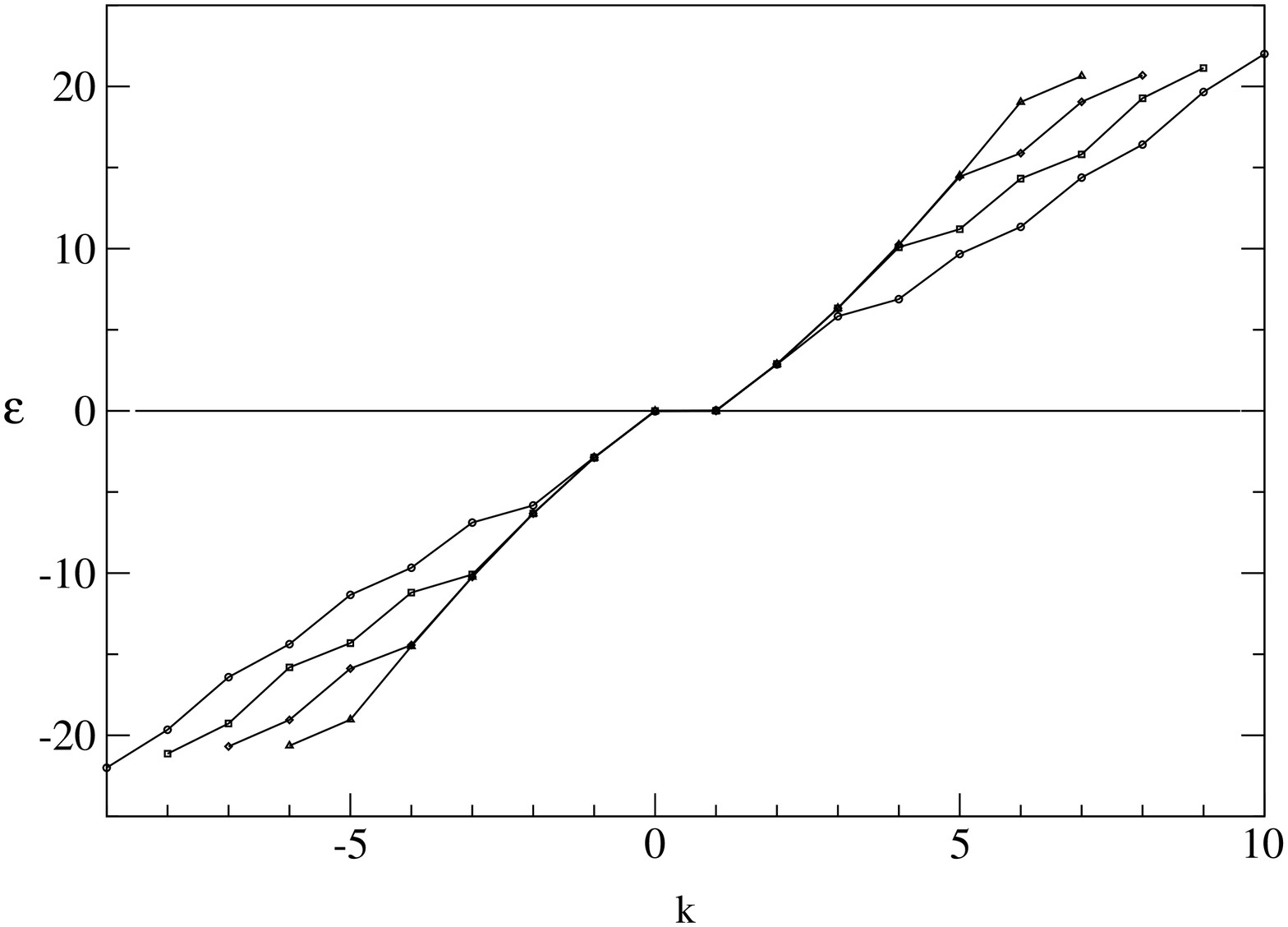,width=7cm,angle=-90}}
\vspace{1mm}
\caption{Low-lying single-particle eigenvalues for a subsystem of L=50
sites with $\it{one}$ boundary defect.\\
 Left : Weak defects, $t=1; 10^{-1}; 10^{-2}; 10^{-3}; 10^{-4}$ 
 (from bottom to top in the right part of the figure). \\
 Right : Strong defects, $t=10; 10^2; 10^3; 10^4$ (also from bottom to top).
 The lines are guides for the eye. \\
The numbering is such that positive $\varepsilon_k$ 
have positive values of $k$. }
\label{fig1.eps}
\end{figure}

localized states 
below and above the band then exhaust the local Hilbert space at the two bond sites and 
effectively cut the system. Thus $C_{1j} = 0$ for $j > 1$ and the diagonal term $C_{11} = 1/2$ gives 
one eigenvalue zero. The other one appears because the remaining part of the subsystem has
an odd number ($L-1$) of sites. It is absent, if the full subsystem has $L+1$ sites. Then the spectrum 
for $t \gg 1$ is exactly the same as for $L$ sites and $1/t \ll 1$ up to the remaining zero eigenvalue.

\begin{figure}
\centerline{\psfig{file=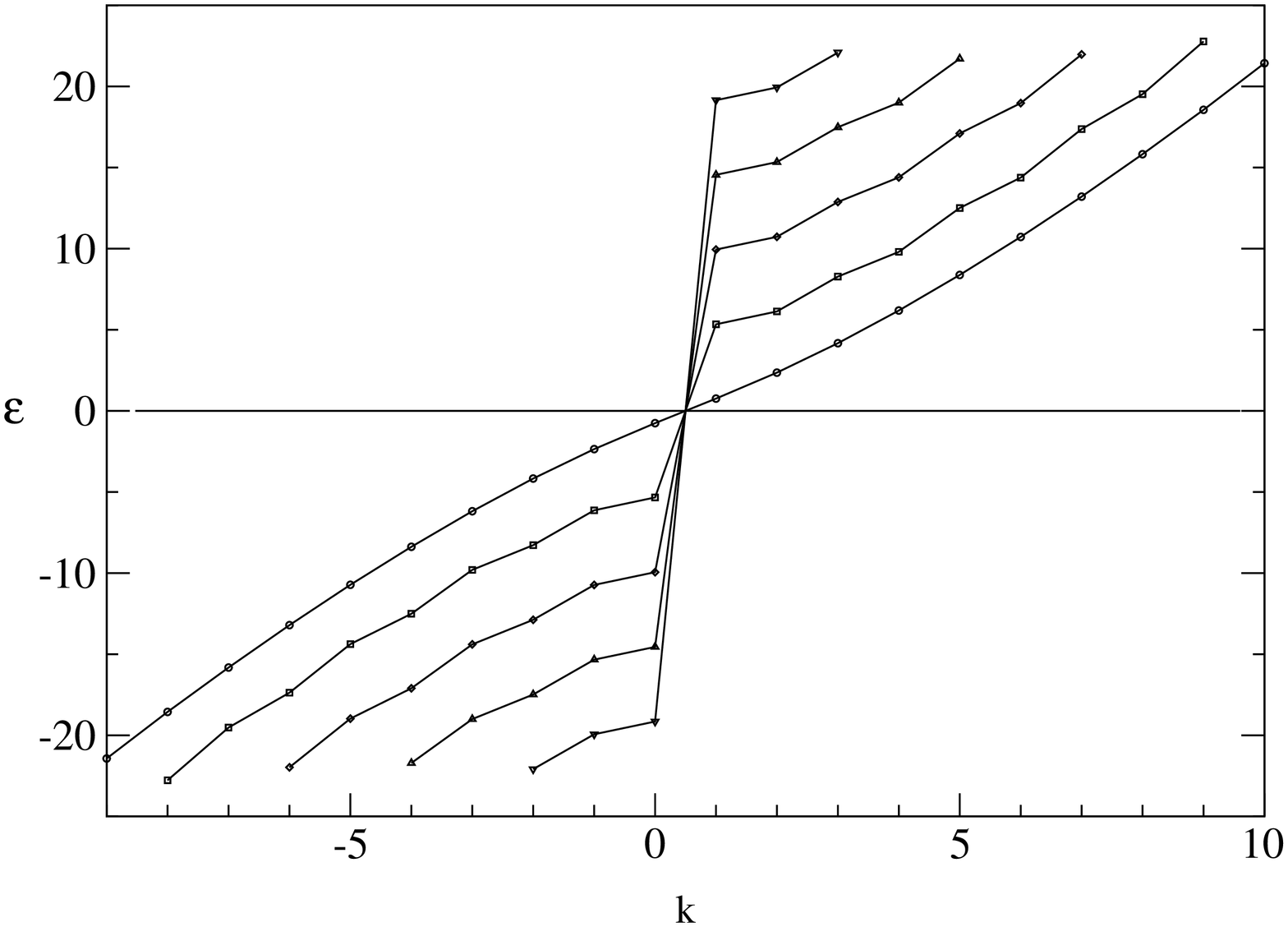,width=7cm,angle=-90}
\psfig{file=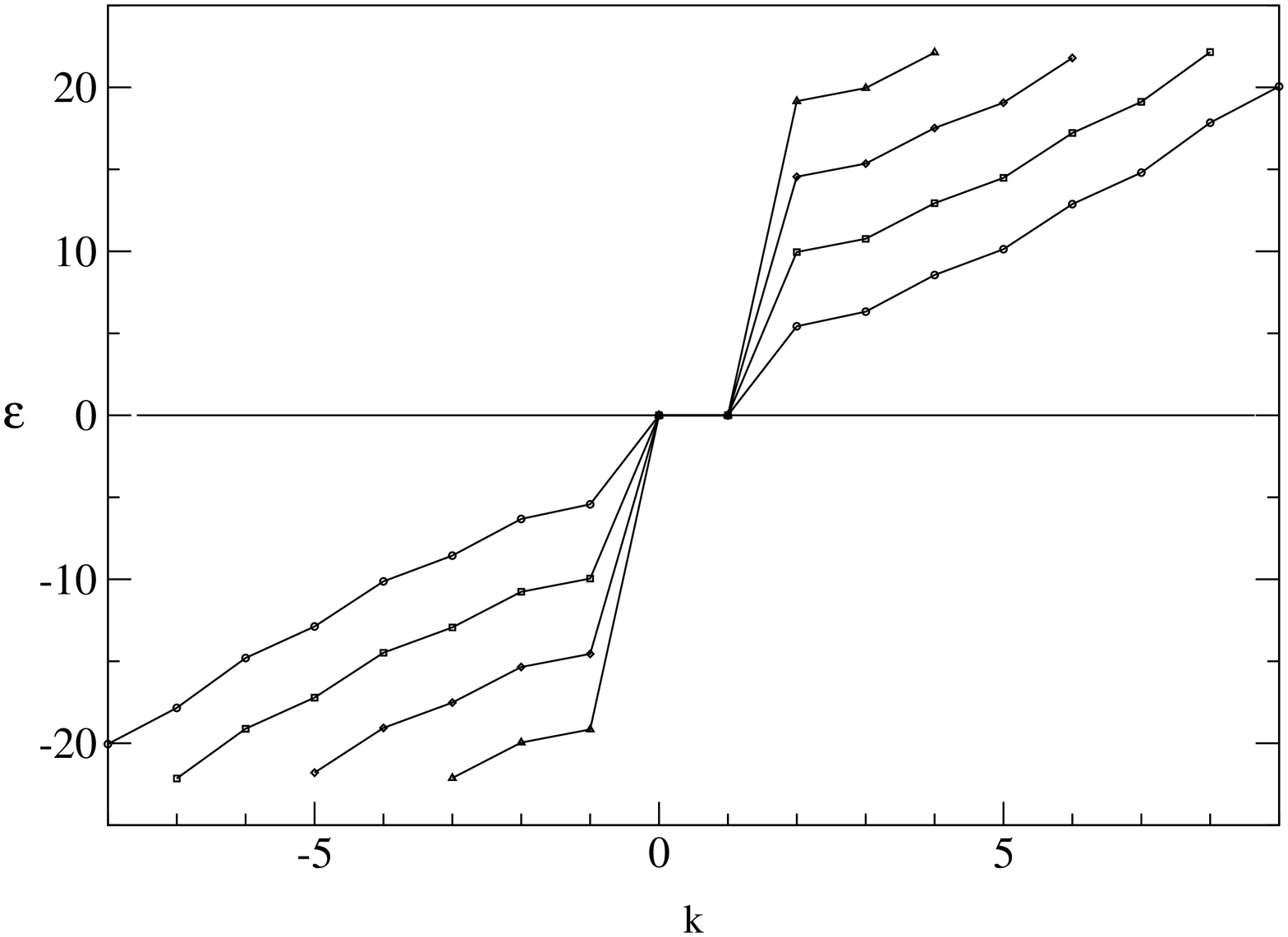,width=7cm,angle=-90}}
\vspace{1mm}
\caption{Low-lying single-particle eigenvalues for a subsystem of $L=50$
sites with $\it{two}$ boundary defects. \\Left : weak bond defects. Right : strong bond defects.
The parameters are the same as in Fig. 1.}
\label{fig2.eps}
\end{figure}

Spectra for two bond defects are shown in Fig. 2. These were calculated by finding
the eigenfunctions of $\hat{H}$ for a total system of size $N=1000$ numerically. This
introduces small finite-size effects but these are not visible on the scale of the figure.
There 
is a shift of the curves as for a single defect, but now it affects $\it{all}$ eigenvalues
if the defects are weak. Thus a gap is opened in the spectrum. For $t \rightarrow 0$  
all $\varepsilon_k$ diverge. This is reflects the fact that the 
subsystem is, in this limit, completely decoupled from the rest and the ground state $|0>$ 
becomes a product state. Then $\rho$ has one eigenvalue 1 if all single-particle
levels are unoccupied and all other eigenvalues vanish. If the defects are strong,
two $\varepsilon_k$ go to zero while the remaining eigenvalues again diverge. The
zero eigenvalues are related to the localized states at the two bonds which belong
to the subsystem as well as to the environment. The ground state $|0>$ then has four 
components which differ by the location of the electrons on the two bonds and the four
resulting eigenvalues 1/4 of $\rho$ are produced by the two vanishing $\varepsilon_k$.\\

For comparison, the spectra for two site defects are shown in Fig. 2. These defects 
destroy the particle-hole symmetry of the problem and therefore the spectrum of the 
$\varepsilon_k$ is in general no longer symmetric with respect to zero. Apart from that,
it shows the same features as for two bond defects. In particular, also the site defects
cut off the subsystem if the site energy $\Delta$ becomes large and thus all 
$\varepsilon_k$ except two diverge for $\Delta\rightarrow \infty$. In this limit, the reflection
symmetry of the spectrum is also restored.

\begin{figure}
\centerline{\psfig{file=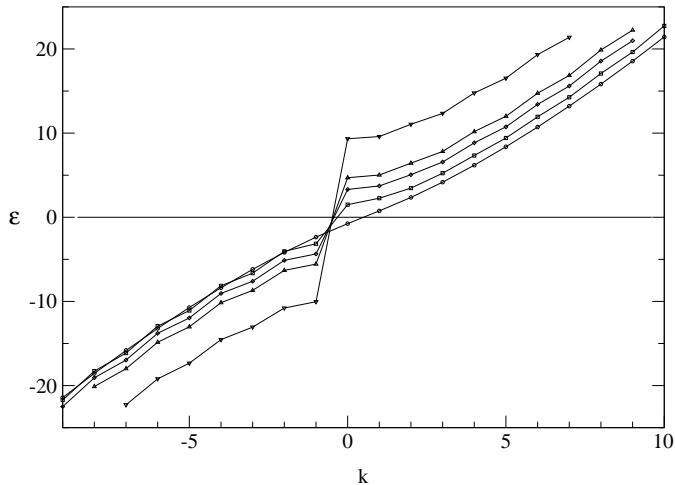,width=8cm,angle=-90}}
\vspace{1mm}
\caption{Low-lying single-particle eigenvalues for a subsystem of $L=50$
sites with two $\it{site}$ defects at the boundaries. The site energies are 
$\Delta= 0; 2; 5; 10; 100$ from bottom to top in the right part of the figure.}
\label{fig3.eps}
\end{figure}

 \section{Entanglement entropy}
 
 Using the values of the $\varepsilon_k$ or the $\zeta_k$, one can obtain the
 the entanglement entropy from (\ref{eqn:entro2a}),(\ref{eqn:entro2b}). Results 
 for single bond defects are shown in Fig. 4 for system sizes between 
 $L=20$ and $L=100$.

  The plot on the right hand side shows that $S$ varies logarithmically with
  $L$ in all cases. There is no indication of a $\ln^2 L$-term as found in 
  \cite{Levine04}. There is only a small variation of the slope with $L$ and
 its asymptotic value can be determined very well by an extrapolation in $1/L$.
 It is remarkable that one can see this logarithmic behaviour so well, since for 
 all sizes considered here only about 20 single-particle eigenvalues contribute 
 if the accuracy is set to $10^{-10}$. Thus one is far away from a limit in which 
 the eigenvalue spectrum becomes dense and where the logarithmic dependence 
 could result from the density of states.

 \begin{figure}
\centerline{\psfig{file=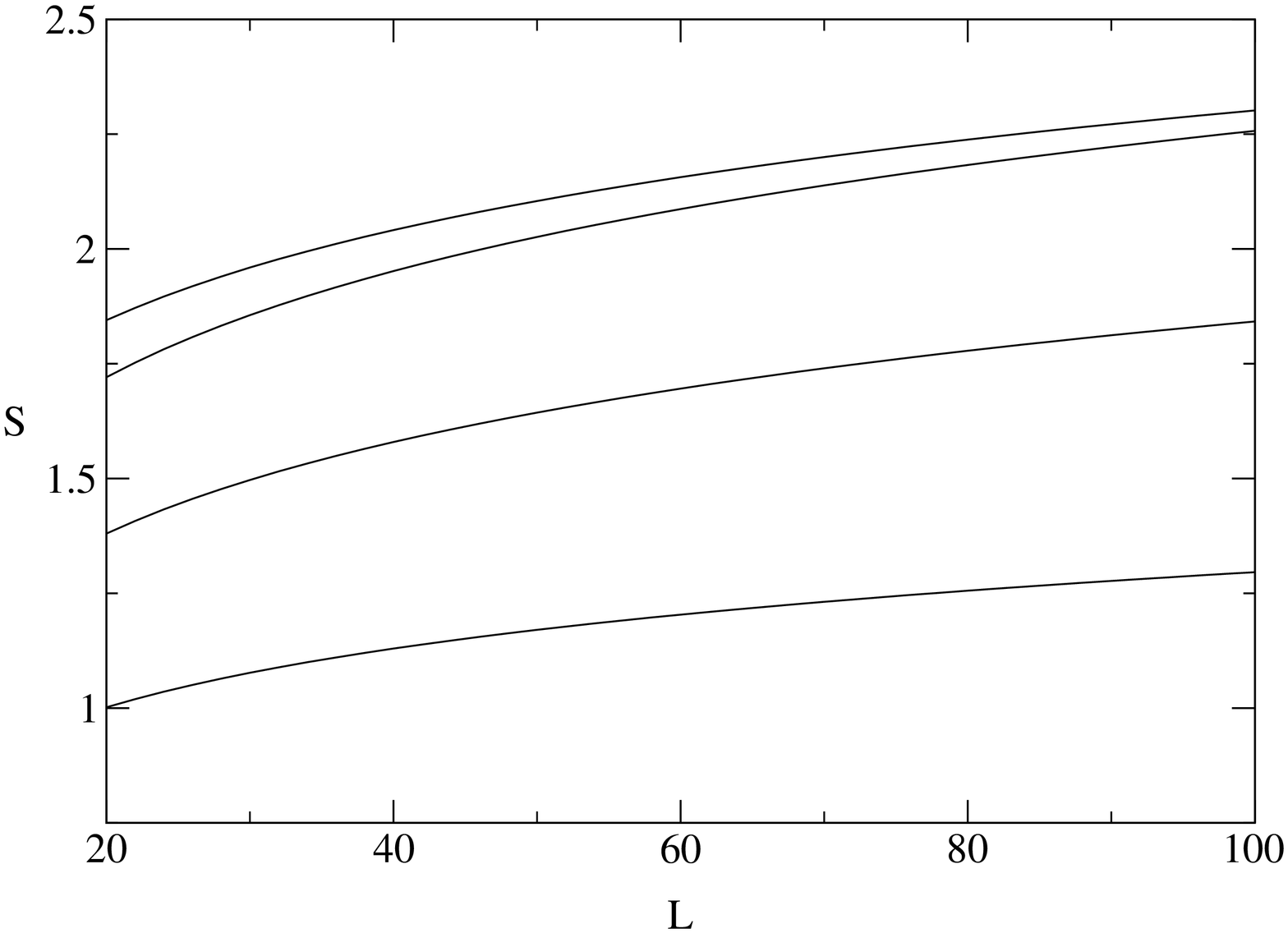,width=6.5cm,angle=-90}
 \psfig{file=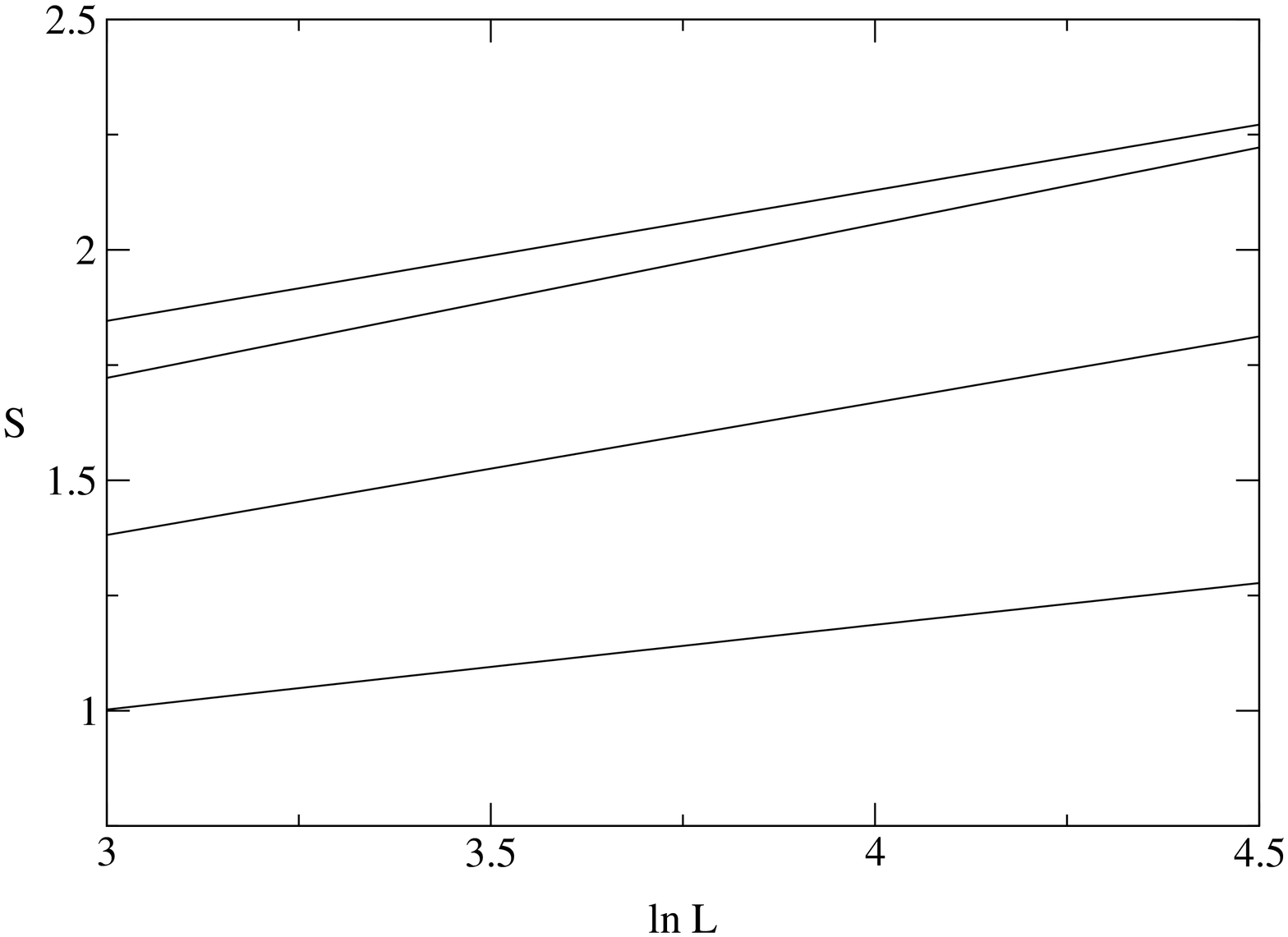,width=6.5cm,angle=-90}}
\vspace{1mm}
\caption{Entanglement entropy as a function of $L$ for one bond defect. 
From bottom to top: t=0.1; 0.5; 1.0; 2.0}
\label{fig4.eps}
\end{figure}

 Thus we find that the entanglement entropy has the same form (\ref{eqn:entro1})
 as for the homogeneous system, but with an effective value $c_{eff}$ which 
 depends on the strength of the defect. The same holds for two bond defects 
 and also for site defects (see below). We first discuss the case of one bond 
 defect further.\\

  The asymptotic value of $c_{eff}$, determined by extrapolating the data from
 $L=20-100$, is shown in Fig. 5 as a function of the defect strength. 
 
 \begin{figure}
\centerline{\psfig{file=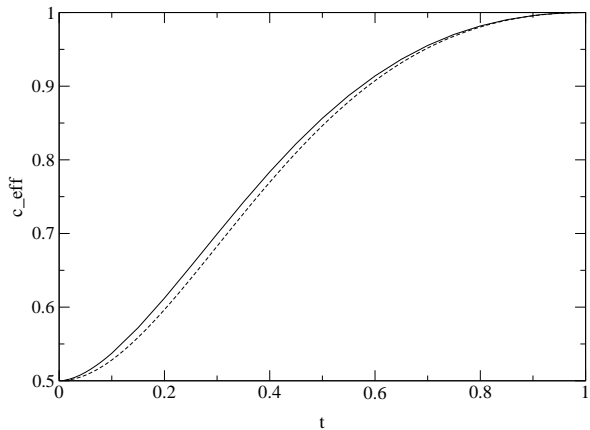,width=7.0cm,angle=-90}}
\vspace{1mm}
\caption{Effective central charge as a function of the bond strength for one bond defect.
 Full line : numerical result, dashed line : approximation Eqn. (\ref{eqn:cfit}).}
\label{fig5.eps}
\end{figure}

 One sees that
 it varies smoothly between the limits $c_{eff}=1/2$  for a subsystem with a free end
 and  $c_{eff}=1$ for a homogeneous system. The figure shows $c_{eff}$ only for $t < 1$
 because one finds that $c_{eff}$ is symmetric under $t \rightarrow 1/t$.

 One can try to find a simple analytic function which fits the numerical curve in Fig. 5.
 Since the defect is characterized by the scattering phases, one could expect that
 $c_{eff}$ depends on these. Because one is dealing with an asymptotic property, the
 phase shifts at the Fermi surface should enter. These are given by
 
  \be
   \delta_{\pm} = \pm \,\bigl[\,2 \,\hbox{arctg}\,(t) - \frac {\pi} {2}\, \bigr]
   \label{eqn:phases}
  \ee
  
 for functions symmetric and antisymmetric about the defect, respectively. One is then
 lead to the expression
 
  \be
   c_{eff} = \frac {1} {2} + \frac {1} {2}\,
   \bigl[\, \frac {4} {\pi} \,\hbox{arctg}(t)\; \frac {4} {\pi} \,\hbox{arctg}(1/t)\,\bigr]^2
   \label{eqn:cfit}
  \ee
  
 The expression in the brackets also appears (as an exponent) if one calculates the time 
 autocorrelation function of a spin in an XX spin chain with a bond defect 
 \cite{Peschel/Schotte84}. The function 
 (\ref{eqn:cfit}) has the correct limits for $t=0$ and $t=1$ and is symmetric under
 $t \rightarrow 1/t$. As can be seen from the figure, it approximates the numerical data 
 quite well. The deviations are less than $10^{-2}$, but the agreement is not perfect.
 Especially for $t \rightarrow 0$, the data do not seem to vary quadratically in $t$ but
 with a power 1.8. Replacing the power 2 in (\ref{eqn:cfit}) by this value improves the
 fit over the whole intervall substantially. The deviations become less than 
 $10^{-3}$ and are almost invisible on the scale of the figure.\\
 
 The two terms in (\ref{eqn:cfit}) can be viewed as the contributions of the unperturbed
 and the perturbed interface, respectively.
 One can understand this result from the form of the single-particle spectrum. In the
 previous section it was found that it consists of two pieces.  If one assumes that in the 
 limit $L \rightarrow \infty$  the levels in these two branches become equidistant and
 dense such that (for positive $\varepsilon_k$)
 \be
   \varepsilon_k= \left \{ \begin {array} {c} 2k\;a/\ln L ,\;\;\; 
                             \varepsilon_k < \alpha  \; \;\;  \\[0.3cm]
          k\; a/\ln L , \;\;\;  \varepsilon_k > \alpha \;\;\;  \; \;\; \end{array} \right.
  \label{eqn:spec1}
 \ee
 where $\alpha = \alpha(t)$ depends on $t$ such that $\alpha = 0$ for $t=1$ and 
 $\alpha = \infty$ for $t=0$ and $a$ is a constant, the entanglement entropy
 can be written as
 \be
  S =  2 \ln L \, \left [ \,\frac {1} {2a} \int_{0}^{\alpha} d\varepsilon \; s(\varepsilon) +
                        \frac {1} {a}  \int_{\alpha}^{\infty} d\varepsilon \;s(\varepsilon) \right ]
   \label{eqn:entro2}
  \ee
  
  Here $s(\varepsilon)$ is given by the bracket in (\ref{eqn:entro2b}) 
  and the factor of 2 in front takes care of the
  negative levels. Denoting the second integral by $G(\alpha)$, one obtains
  \be
    S = 2 \ln L \; \bigl[\, \frac {1} {2} (G(0) - G(\alpha)) + G(\alpha) \,\bigr]
        =  \ln L \; \bigl[\,G(0) + G(\alpha)\,\bigr]
   \label{eqn:entro3}
   \ee
  Thus each interface brings its own amplitude for the logarithm : the unperturbed
  one $G(0)$ and the perturbed one $G(\alpha)$.
  For the homogeneous system the bracket must have the value 1/3, from which
  $G(0)=1/6$ follows. Therefore (\ref{eqn:entro3}) gives for $c_{eff}$
  \be
    c_{eff} = \frac {1} {2} + 3\; G(\alpha)
   \label{eqn:ceff1}
  \ee
  If the exact dependence of $\alpha$ on $t$ were known, one could evaluate the
  integral $G(\alpha)$ and determine $c_{eff}$ in this way. The oscillations of the
  $\varepsilon_k$ which are neglected in (\ref{eqn:spec1}) could also be included.\\
  
  The discussion so far has dealt with one defect. However, one can apply it also to
  the case of two defects. The results of Section III show that then the lowest
  eigenvalues are just missing and one also finds that the gap in the spectrum for
  two defects  
  corresponds roughly to the bending point in the spectrum for one defect. 
  Therefore one has only the second region in (\ref{eqn:spec1}) and finds instead of
  (\ref{eqn:entro3})
  \be
    S =   \ln L \; [2 \;G(\alpha)]
   \label{eqn:entro4}
   \ee
  from which the value of $c_{eff}$ can be read off. This also gives the relation
  \be 
    c^{(2)}_{eff} = 2c^{(1)}_{eff} - 1
    \label{eqn:ceff2}
  \ee
  between the effective c-values for two defects and for one defect. Due to the
  lower accuracy of the calculations for two defects, this relation could not be
  checked precisely, but the deviations are less than some percent and vanish near
  $t=0$ and $t=1$.\\
  
 Results for the case of one site defect are shown in Fig. 6. Here the entanglement
 entropy is symmetric under $\Delta \rightarrow -\Delta$ and only positive values
 of $\Delta$ have to be considered. One sees that $c_{eff}$ decreases again from
 the value 1 to 1/2 as the perturbation becomes stronger, because
 the defect cuts the chain in the limit $\Delta \rightarrow \infty$ and thus acts
 like a bond with $t=0$ or $t=\infty$. 
    
 \begin{figure}
\centerline{\psfig{file=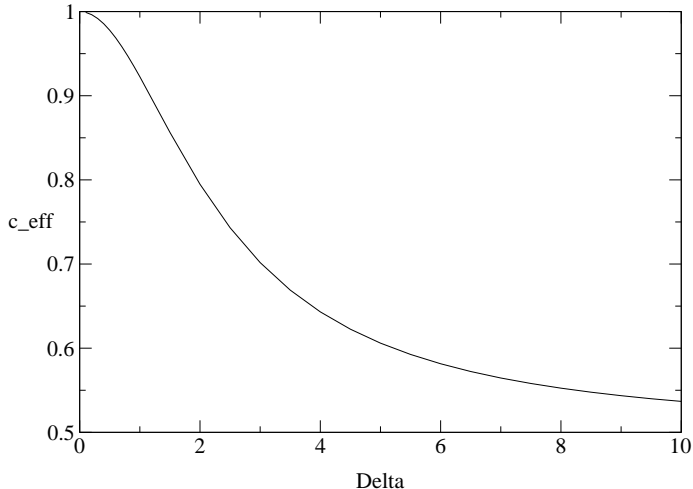,width=8cm,angle=-90}}
\vspace{1mm}
\caption{Effective central charge as a function of the site energy $\Delta$ for one site defect.
 The values were obtained by extrapolation in $1/L$}
\label{fig6.eps}
\end{figure}

  Finally, we present results for the constant $k$ in the entropy, see (\ref{eqn:entro1}).
  They were obtained by extrapolating $S(L)-(c_{eff}/3) \; ln L$ from the values $L=20-100$.
  As Fig. 7 shows, this quantity $k$ is not symmetric under $t \rightarrow 1/t$ for one bond
  defect. Rather it increases from about 0.5 for $t=0$ to about 1.2 for large
  values of $t$, with a small initial dip (which depends sensitively on $c_{eff}$).
   At $t=1$ we find a value $k=0.726$ which is in
  agreement with the findings in \cite{Latorre03} but differs slightly from the value
  obtained from the asymptotic analysis in \cite{Jin04}. The limiting value for 
  $t \rightarrow \infty$ is found to differ from the value at $t=0$ by $\ln 2$. If one considers 
  systems with an odd number of sites $L+1$ for $t \gg 1$, this also holds and follows in this
  case directly from the identity of the single-particle spectra up to one zero eigenvalue 
  (which contributes $\ln 2$) noted in Section III. 
  The relatively large value of $k$ can make the entropy with a defect larger
  than without one if the subsystem is small enough (compare Fig. 4). However, in the limit 
  $L \rightarrow \infty$ the effect of the reduced $c_{eff}$ always dominates and leads
  to a reduction of the entanglement.
  
 \begin{figure}
\centerline{\psfig{file=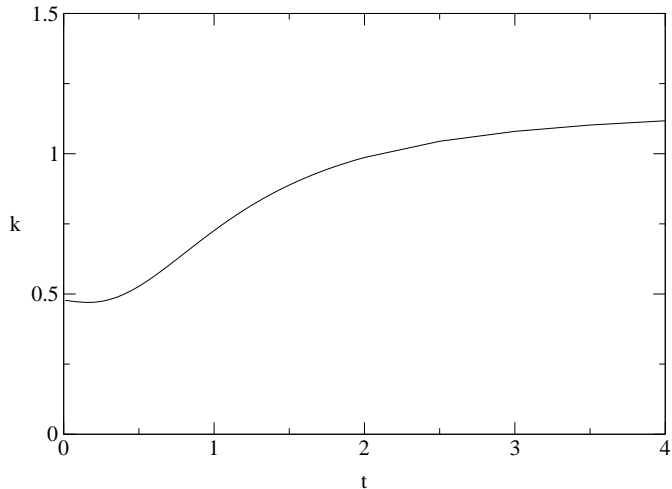,width=8cm,angle=-90}}
\vspace{1mm}
\caption{Constant $k$ in the entanglement entropy as a function of the bond strength
 $t$ for one bond defect. The values were obtained by extrapolation in $1/L$}
\label{fig7.eps}
\end{figure}
  
\section{Conclusion}

We have investigated  a chain of free electrons in its ground state at half filling. 
The entanglement properties were calculated for the case of one or
two defects located at the contact points between the subsystem and the rest of the
chain. The density-matrix spectra were seen to differ in a characteristic way from
those for a homogeneous system. Nevertheless, the logarithmic dependence of 
the entanglement entropy on the size persists, only the amplitude changes 
and becomes dependent on the defect strength.
The mechanism could be described by a simple model based on the features of
the single-particle spectra. A simple approximate formula for the effective central
charge was also given. Although we focussed on bond defects, the same 
features are found for site defects. The logarithmic law means that the system remains 
critical, as expected for a localized perturbation. In the two-dimensional system
associated with the quantum chain, the defect becomes a line as noted and used
in \cite{Levine04}.
Such a two-dimensional system with a cut (because one considers the reduced density
matrix) and one or two straight defect lines perpendicular to this cut can be mapped 
conformally to a strip \cite{Peschel04} but the defect lines are bent by 
the mapping. For a single defect, the line runs only in the left or right half of the strip,
coming from one point on the boundary and going back to another one. 
For two defects, the two lines are attached to different edges and touch in 
the center of the strip. The more complicated density-matrix spectra found here must 
be related to this feature. It would be interesting to derive them analytically. Also
one wonders if the asymptotic formulae used in \cite{Jin04,Keating/Mezzadri04} to 
calculate $S$ can be generalized to the present case.\\

$\it{Acknowledgement}$ The author thanks K.D. Schotte for discussions and help with
the figures.

\end{document}